\documentstyle[12pt,epsf]{article}

\begin{document}

\begin{flushright}
hep-ph/0207296 \\
\end{flushright}
\vspace*{1.5cm}

\begin{center}
{ \bf
THE BFKL POMERON WITHIN PHYSICAL RENORMALIZATION
SCHEMES AND SCALES} \footnote{Proc. XXXV PNPI Winter School, Repino, 
St.Petersburg, Russia, 19--25 February, 2001, \\
eds: V. A. Gordeev {\it et al.}, pp. 259-278 (St.Petersburg, 2001)  

} \\
\begin{center}
{ \large Victor~S.~Fadin${}^{\dagger}$,
Victor~T.~Kim${}^{\ddagger }$, \\
 Lev~N.~Lipatov${}^{\ddagger}$
{\rm and}
Grigorii~B.~Pivovarov${}^{\S}$ } \\ 
\end{center}
\begin{center}
${}^\dagger$  Budker Institute for Nuclear Physics,
Novosibirsk 630090, Russia \\
${}^\ddagger$ St.  Petersburg Nuclear Physics Institute,
Gatchina 188300, Russia \\
${}^\S$  Institute for Nuclear Research, Moscow 117312, Russia
\end{center}

A b s t r a c t
\end{center}

In this lecture  the  next-to-leading order (NLO) corrections to the
QCD Pomeron intercept obtained from the Balitsky-Fadin-Kuraev-Lipatov
(BFKL) equation are discussed.
It is shown that the BFKL Pomeron intercept when evaluated
in non-Abelian physical renormalization schemes with
Brodsky-Le\-pa\-ge-Mackenzie (BLM) optimal scale
setting does not exhibit the serious problems encountered in the
modified minimal subtraction ($\overline{MS}$) scheme.
The results obtained provide an opportunity for applications of
the NLO BFKL resummation to high-energy phenomenology.
 One of such applications
for virtual gamma-gamma total cross section shows a good
agreement
with preliminary data at CERN LEP.

\newpage

\section{Motivation}

The discovery of rapidly increasing structure functions in deep
inelastic
scattering (DIS) at HERA \cite{HERA} at small-$x$ is in agreement
with the
expectations of the QCD high-energy limit.
The Balitsky-Fadin-Kuraev-Lipatov (BFKL) \cite{BFKL,BL78}
resummation of energy
logarithms is anticipated to be an important tool for exploring this
limit.
The leading order (LO) BFKL calculations \cite{BFKL} predict a
steep
rise of QCD cross sections.
Namely, the highest eigenvalue, $\omega^{max}$, of the BFKL
equation \cite{BFKL} is
related to the intercept of the Pomeron which in turn governs
the high-energy asymptotics of the cross sections: $\sigma \sim
s^{\alpha_{I \negthinspace P}-1} = s^{\omega^{max}}$.
The BFKL Pomeron intercept in the LO turns out to be rather large:
$\alpha_{I \negthinspace P} - 1 =\omega_L^{max} =
12 \, \ln2 \, ( \alpha_S/\pi )  \simeq 0.55 $ for
$\alpha_S=0.2$; hence, it is very important to know the next-to-
leading order
(NLO) corrections.
In addition, the LO BFKL calculations have restricted
phenomenological
applications because, {\it e.g.}, the running  of the QCD coupling
constant $\alpha_S$ is not included and the kinematic range of
validity of LO BFKL is not known.

Recently the  NLO corrections to the BFKL resummation
of energy logarithms were calculated; see Refs. \cite{FL,CC98} and
references
therein. The NLO corrections \cite{FL,CC98} to the highest
eigenvalue
of the BFKL equation turn out to be negative and even larger
than the LO contribution for $\alpha_S > 0.157$. At such
circumstances
the phenomenological significance of the NLO BFKL calculations
seems to be
rather obscure.

However, one should stress that the NLO calculations,
as any finite-order perturbative results, contain
both  renormalization scheme and renormalization scale
ambiguities.
The NLO BFKL calculations \cite{FL,CC98} were
performed by employing the modified minimal subtraction scheme
($\overline{\mbox{MS}}$) \cite{Bar78} to regulate the ultraviolet
divergences with arbitrary scale setting.

In this work we
consider the NLO BFKL resummation of energy logarithms
\cite{FL,CC98} in physical renormalization schemes in order to
study the
renormalization scheme dependence. To resolve the
renormalization scale
ambiguity we utilize  Brodsky-Lepage-Mackenzie (BLM) optimal
scale
setting \cite{BLM}.
We show that the reliability of QCD predictions for the intercept of
the
BFKL Pomeron at NLO when evaluated using  BLM scale setting
within non-Abelian
physical schemes, such as the momentum space
subtraction (MOM) scheme \cite{Cel79,Pas80}
or the $\Upsilon$-scheme based on  $\Upsilon \rightarrow ggg$
decay,
is significantly improved as compared to the $\overline{\mbox{MS}}$-
scheme.
This provides a basis for applications of NLO BFKL resummation to
high-energy phenomenology. Certain aspects of this work were
presented in Ref. \cite{BFKLP}.

\section{BFKL in Physical Renormalization Schemes}

We begin with the representation of the $\overline{\mbox{MS}}$-
result
of NLO BFKL \cite{FL,CC98} in physical renormalization schemes.
Although the $\overline{\mbox{MS}}$-scheme is somewhat artificial
and it
lacks a clear physical picture, it
can serve as a convenient intermediate renormalization
scheme. The eigenvalue of the NLO BFKL equation  at transferred
momentum
squared $t=0$ in the $\overline{\mbox{MS}}$-scheme
\cite{FL,CC98} can be represented as the action of
the NLO BFKL kernel (averaged over azimuthal angle) on the LO
eigenfunctions $(Q_{2}^{2}/Q_{1}^{2})^{-1/2+i\nu }$  \cite{FL}:
\begin{eqnarray}
\omega _{\overline{MS}}(Q_{1}^{2},\nu ) &=&\int d^{2}Q_{2}\,\,\,
K_{\overline{MS}}(\vec{Q_{1}},\vec{Q_{2}})
\left( \frac{Q_{2}^{2}}{Q_{1}^{2}}
\right) ^{-\frac{1}{2}+i\nu }=  \nonumber \\
&=&  N_C \chi_{L}(\nu ) \frac{\alpha_{\overline{MS}}(Q_{1}^{2})}{\pi }
\Biggl[ 1 + r_{\overline{MS}}(\nu )
\frac{\alpha_{\overline{MS}}(Q_{1}^{2})}{\pi } \Biggr] ,
\label{kernelact}
\end{eqnarray}
where
\[
\chi _{L}(\nu )\ \ =2\psi (1)-\psi (1/2+i\nu)-\psi (1/2-i\nu)
\]
is the function related with the LO eigenvalue,
$\psi =\Gamma ^{\prime}/\Gamma $ denotes the Euler $\psi $-
function, the
$\nu $-variable is a conformal weight parameter \cite{Lipatov97},
$N_C$ is
the number of colors, and $Q_{1,2}$ are the virtualities
of the reggeized gluons.

The calculations of Refs. \cite{FL,CC98} allow us
to decompose the NLO coefficient $r_{\overline{MS}}$ of
Eq. (\ref{kernelact}) into $\beta$-dependent and
the conformal ($\beta$-independent) parts:
\begin{equation}
r_{ \overline{MS}} (\nu)\ =\ r_{\overline{MS}}^{\beta}(\nu)
+ r_{ \overline{MS}}^{conf} (\nu)\ ,
\label{evnl}
\end{equation}
where
\begin{equation}
r_{ \overline{MS}}^{\beta}(\nu)\  =\ - \frac{\beta_0}{4} \Biggl[
\frac{1}{2} \chi_{L}(\nu ) - \frac{5}{3} \Biggr]
\label{chimsbeta}
\end{equation}
and
\begin{eqnarray}
r_{\overline{MS}}^{conf} (\nu)
& = &  - \frac{N_C}{4 \chi_{L}(\nu )} \left[
\frac{\pi^2 \sinh (\pi \nu )}
{2 \nu \cosh^2 (\pi \nu )}
\left( 3 + \left( 1+ \frac{N_F}{N_C^3} \right) \right.
\frac{11 + 12 \nu^2 }
{16 (1 + \nu^2 )} \right) - \chi_{L}^{\prime \prime }(\nu )
 \nonumber \\
& &  +\  \left.\frac{\pi^2-4}{3} \chi_{L}(\nu ) -
\frac{\pi^3}{\cosh(\pi \nu )} - 6\zeta(3) + 4\varphi(\nu) \right]
\label{chimsconf}
\end{eqnarray}
with
\begin{equation}
\varphi (\nu ) = 2 \int_0^1 dx \frac{\cos(\nu \ln(x))}{(1+x) \sqrt{x}}
\Biggl[ \frac{\pi^2}{6} - {\mathrm Li}_2 (x) \Biggr], \;
{\mathrm Li}_2 (x) = - \int_0^x dt \frac{\ln (1-t)}{t}.
\end{equation}
Here $\beta_0 = (11/3)N_C - (2/3) N_{F}$ is the leading coefficient
of the QCD $\beta$-function, $N_F$ is the number of flavors,
$\zeta(n)$ stands for the Riemann zeta-function,
${\mathrm Li}_2 (x)$
is the Euler dilogarithm (Spence-function).
In Eq. (\ref{chimsconf}) $N_F$  denotes flavor number of
the Abelian part of the
$gg \rightarrow q\overline{q}$ process contribution.
The Abelian part is not associated with the running of the
coupling \cite{BH} and is consistent with the correspondent QED
result for the $\gamma^{\ast} \gamma^{\ast} \rightarrow e^+e^- $
cross section \cite{GLF}.

The $\beta$-dependent NLO coefficient  $r_{ \overline{MS}}^{\beta}
(\nu)$,
which is related to the running of the coupling,
receives contributions from the gluon reggeization diagrams,
from the virtual part of the one-gluon emission, from the real
two-gluon emission, and from the non-Abelian part \cite{BH} of the
$gg \rightarrow q\overline{q}$ process.
There is an omitted term in $r_{\overline{MS}}^{\beta} (\nu )$
proportional to $\chi^{\prime}_L(\nu)$ which originates
from the asymmetric treatment of $Q_1$ and $Q_2$,
it  can be removed by the redefinition of the LO
eigenfunctions \cite{FL}.

The NLO BFKL Pomeron intercept then  reads for $N_C=3$
\cite{FL}:
\begin{eqnarray}
&&\alpha_{I \negthinspace P}^{\overline{MS}} - 1  =
\omega_{\overline{MS}}(Q^2,0) =
12 \, \ln2 \, \frac{ \alpha_{\overline{MS}}(Q^2)}{\pi} \biggl[
1+r_{\overline{MS}}(0)
\frac{\alpha_{\overline{MS}}(Q^2)}{\pi} \biggr]  ,\\
&& r_{\overline{MS}}\ (0)\ \simeq\ -20.12-0.1020
N_F+0.06692\beta_0\ ,
\label{rms0}\\
&& r_{\overline{MS}}(0)_{\vert N_F =4}\ \simeq\ -19.99\ .
\nonumber
\end{eqnarray}

Physical renormalization schemes provide small and physically
meaningful
perturbative coefficients by incorporating large corrections into the
definition of the coupling constant.
One of the most popular physical schemes
is MOM-scheme
\cite{Cel79,Pas80}, based on renormalization of the triple-gluon
vertex
at some symmetric off-shell momentum. However, in the MOM-
scheme
the coupling constant is gauge-dependent already in the LO, and
there
are rather cumbersome technical difficulties. These
difficulties can be avoided by performing calculations in the
intermediate
$\overline{\mbox{MS}}$-scheme, and then by making the transition
to some
physical scheme by a finite renormalization \cite{Cel79}.
In order to eliminate the dependence on the gauge choice and other
theoretical
conventions, one can consider renormalization
schemes based on physical processes \cite{BLM}, {\it e.g.},
V-scheme based on heavy quark potential.
Alternatively, one can introduce a physical scheme based on the
$\Upsilon \rightarrow ggg$ decay using the NLO calculations of
Ref. \cite{ML}.

A finite renormalization due to the change of scheme can be
accomplished by
a transformation of the QCD coupling \cite{Cel79}:
\begin{equation}
\alpha_S\ \rightarrow\ \alpha_{S} \biggl[ 1 + T \frac{\alpha_{S}}{\pi}
\biggr] ,
\end{equation}
where $T$ is some function of $N_C$, $N_F$ and, for the MOM-
scheme, of
a gauge parameter $\xi$. Then the NLO BFKL eigenvalue in
the MOM-scheme can be represented  as follows:
\begin{eqnarray}
\omega_{MOM}(Q^2,\nu)
& = & N_C \chi_{L}(\nu) \frac{\alpha_{MOM} (Q^2)}{\pi} \biggl[ 1 +
r_{MOM}(\nu)\,\, \frac{\alpha_{MOM}(Q^2)} {\pi}\biggr] \, , \\
r_{MOM}(\nu) & = & r_{\overline{MS}} (\nu) + T_{MOM} . \nonumber
\end{eqnarray}
For the transition from the $\overline{\mbox{MS}}$-scheme to the
MOM-scheme the corresponding T-function has the following form
\cite{Cel79}:
\begin{eqnarray}
T_{MOM}& = & T_{MOM}^{conf}+T_{MOM}^{\beta}, \\
T_{MOM}^{conf} &=& \frac{N_C}{8} \biggl[ \frac{17}{2} I +
\xi \frac{3}{2} (I-1) + \xi^2 (1-\frac{1}{3}I) -
\xi^3 \frac{1}{6} \biggr]  , \nonumber \\
T_{MOM}^{\beta} &=& - \frac{ \beta_0}{2} \biggl[ 1 +\frac{2}{3} I
\bigg] ,
\nonumber
\end{eqnarray}
where $I=-2 \int^{1}_{0}dx \ln(x)/[x^2-x+1]\simeq 2.3439$.

Likewise, one can obtain for the V-scheme \cite{BLM}:
\begin{equation}
T_V\  =\ \frac{2}{3} N_C -\frac5{12}\ \beta_0\ ,
\end{equation}
and, by the use of the results of Ref. \cite{ML}, for the $\Upsilon$-
scheme:
\begin{equation}
T_{\Upsilon}\ =\ \frac{6.47}3\ N_C- \frac{2.77}3\ \beta_0\ .
\end{equation}

One can see from Table \ref{fig:1} that
there is no a strong renormalization scheme dependence, though
the problem of a large NLO
BFKL coefficient remains. A large size of the perturbative
corrections leads to a significant renormalization scale ambiguity.

\begin{table}
\caption{Scheme-transition function and the NLO BFKL coefficient
in physical schemes}
\begin{center}
\begin{tabular}{|c|c|c|c|c|}
\hline
\multicolumn{2}{|c|}{Scheme}  & $T=T^{conf}+T^{\beta}$ &
 $r(0)=r^{conf}(0)+r^{\beta}(0)$
& $r(0)$        \\
\multicolumn{2}{|c|}{} & & & $(N_F = 4)$ \\
\hline\hline
M & $\xi=0$ & $7.471 - 1.281 \beta_0$ &
$-12.64 - 0.1020 N_F - 1.214 \beta_0$ & -22.76 \\
 \cline{2-5}
O & $\xi=1$ & $8.247 - 1.281 \beta_0$ &
$-11.87 - 0.1020 N_F - 1.214 \beta_0$ & -21.99 \\
\cline{2-5}
M & $\xi=3$ & $8.790 -1.281 \beta_0$ &
$-11.33 - 0.1020 N_F - 1.214 \beta_0$& -21.44  \\
\hline
\multicolumn{2}{|c|}{V} & $ 2 - 0.4167 \beta_0 $ &
$-18.12 - 0.1020 N_F - 0.3497 \beta_0$ & -21.44 \\ \hline
\multicolumn{2}{|c|}{$\Upsilon$} & $6.47 - 0.923 \beta_0$ &
$-13.6 - 0.102 N_F - 0.856 \beta_0 $ & -21.7  \\
\hline
\end{tabular}
\end{center}
\label{tab:1}
\end{table}

\section{Optimal Renormalization Scale Setting}

The renormalization scale ambiguity problem can be resolved if one
can
optimize the choice of scales and renormalization schemes
according to some
sensible criteria. In the BLM optimal scale
setting \cite{BLM}, the renormalization scales are chosen such that
all vacuum polarization effects from the QCD $\beta$-function are
resummed
into the running couplings. The coefficients of the perturbative
series are
thus identical to the perturbative coefficients of the corresponding
conformally invariant theory with $\beta=0$. The BLM
approach has an important advantage of resumming the large and
strongly
divergent terms in the perturbative QCD series which grow as $n!
[\alpha_S
\beta_0 ]^n$, {\it i.e.}, the infrared renormalons
associated with coupling constant renormalization.
The renormalization scales in the BLM approach are physical in
the sense that they reflect the mean virtuality of the gluon
propagators \cite{BLM}.

The BLM scale setting \cite{BLM} can be applied within any
appropriate physical scheme.
In the present case one can show that within the
V-scheme (or the $\overline{\mbox{MS}}$-scheme)
the BLM procedure does not change significantly the value of
the NLO coefficient $r(\nu)$.
This can be understood since the V-scheme as well as
$\overline{\mbox{MS}}$-scheme are primarily adjusted to
the case when, in the LO, there are dominant QED (Abelian) type
contributions, whereas, in the BFKL case,  the LO
gluon-gluon (non-Abelian) interactions are important.

Therefore, from the point of view of BLM scale setting, one
can separate QCD processes into two classes specifying whether
gluons are
involved into the LO or not.
Thus one can expect that in the BFKL case
it is appropriate to use a physical scheme which is adjusted to
non-Abelian interactions in the LO. One can choose the MOM-
scheme
based on  the symmetric triple-gluon
vertex \cite{Cel79,Pas80} or the $\Upsilon$-scheme based on
$\Upsilon \rightarrow ggg$ decay.
The importance of taking into account this circumstance
for vacuum polarization effects can be seen from the
``incorrect'' sign of the $\beta_0$-term for $r_{\overline{MS}}$
in the unphysical $\overline{\mbox{MS}}$-scheme (Eq. (\ref{rms0})).

Adopting  BLM scale setting, the NLO BFKL eigenvalue
in the MOM-scheme is
\begin{equation}
\omega_{BLM}^{MOM}(Q^{2},\nu) =
N_C \chi_{L} (\nu) \frac{\alpha_{MOM}(Q^{MOM \, 2}_{BLM})}{\pi}
\Biggl[1 +
r_{BLM}^{MOM} (\nu) \frac{\alpha_{MOM}(Q^{MOM \,
2}_{BLM})}{\pi} \Biggr] ,
\end{equation}
\begin{equation}
r_{BLM}^{MOM} (\nu) = r_{MOM}^{conf} (\nu) \, .
\end{equation}

The $\beta$-dependent part of the $r_{MOM}(\nu)$ defines the
corresponding BLM optimal scale
\begin{equation}
Q^{MOM \, 2}_{BLM} (\nu) = Q^2 \exp
\Biggl[ - \frac{4 r_{MOM}^{\beta}(\nu)}{\beta_0} \Biggr]
= Q^2 \exp \Biggl[ \frac
{1}{2}\chi_L (\nu) - \frac{5}{3} + 2 \biggl(1+\frac{2}{3} I \biggr) \Biggr].
\label{qblm}
\end{equation}
Taking into account the fact that $\chi_L(\nu) \rightarrow -2 \ln (\nu)
$ at
$\nu \rightarrow \infty $, one obtains at large $\nu$
\begin{equation}
Q^{MOM\,2}_{BLM}(\nu)\ =\ Q^2 \frac1\nu \exp\left[2\biggl(1 +
\frac23\ I \biggr) - \frac53 \right].
\end{equation}

At $\nu=0$ we
have $Q^{MOM \, 2}_{BLM} (0) = Q^2 \bigl( 4 \exp [2(1+2 I /3)-5/3]
\bigr) \simeq
Q^2  \, 127$. Note that $Q^{MOM \, 2}_{BLM}(\nu)$ contains a
large factor,
$\exp [- 4 T_{MOM}^{\beta}/\beta_0 ] = \exp [2(1+2 I /3)] \simeq
168$, which
reflects a large kinematic difference between MOM- and
$\overline{\mbox{MS}}$- schemes \cite{Cel83,BLM}, even in an
Abelian
theory.

\begin{table}
\caption{The NLO- BFKL-Pomeron intercept in the BLM scale
setting
within non-Abelian physical schemes}
\begin{center}
\begin{tabular}{|c|c|c|c|c|c|}
\hline
\multicolumn{2}{|c|}{Scheme} & \multicolumn{1}{|c|}{$r_{BLM}(0)$} &
\multicolumn{3}{|c|}
{$\alpha_{I \negthinspace P}^{BLM} - 1 =\omega_{BLM}(Q^2,0)$} \\
\cline{4-6}
\multicolumn{2}{|c|}{}  & \multicolumn{1}{|c|}{$(N_F = 4)$} &
$Q^2=1$ GeV$^2$ & $Q^2=15$ GeV$^2$ & $Q^2=100$ GeV$^2$
\\
\hline\hline
M & $\xi=0$ & -13.05 & 0.134 & 0.155 & 0.157 \\
 \cline{2-6}
O & $\xi=1$ & -12.28 & 0.152 &  0.167 & 0.166 \\
\cline{2-6}
M & $\xi=3$ & -11.74 & 0.165 & 0.175 & 0.173 \\
\hline
\multicolumn{2}{|c|}{$\Upsilon$} & -14.01 & 0.133  & 0.146 & 0.146
\\
\hline
\end{tabular}
\end{center}
\label{tab:2}
\end{table}

Analogously, one can implement the BLM scale setting in the
$\Upsilon$-scheme (Table \ref{tab:2}).

\begin{figure}[htb]
\begin{center}
\leavevmode
{\epsfxsize=8.5cm\epsfysize=8.5cm\epsfbox{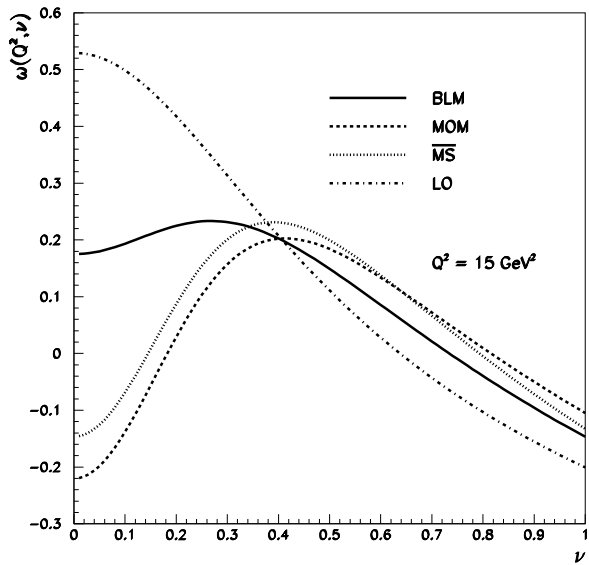}}
\end{center}
\caption[*]{$\nu$-dependence of the NLO BFKL eigenvalue at
$Q^2=15$
GeV$^2$: BLM (in MOM-scheme) -- solid,
MOM-scheme (Yennie gauge: $\xi=3$) -- dashed,
$\overline{\mbox{MS}}$-scheme -- dotted. LO
BFKL ($\alpha_S=0.2$) -- dash-dotted.}
\label{fig:1}
\end{figure}

\begin{figure}[htb]
\begin{center}
\leavevmode
{\epsfxsize=8.5cm\epsfysize=8.5cm\epsfbox{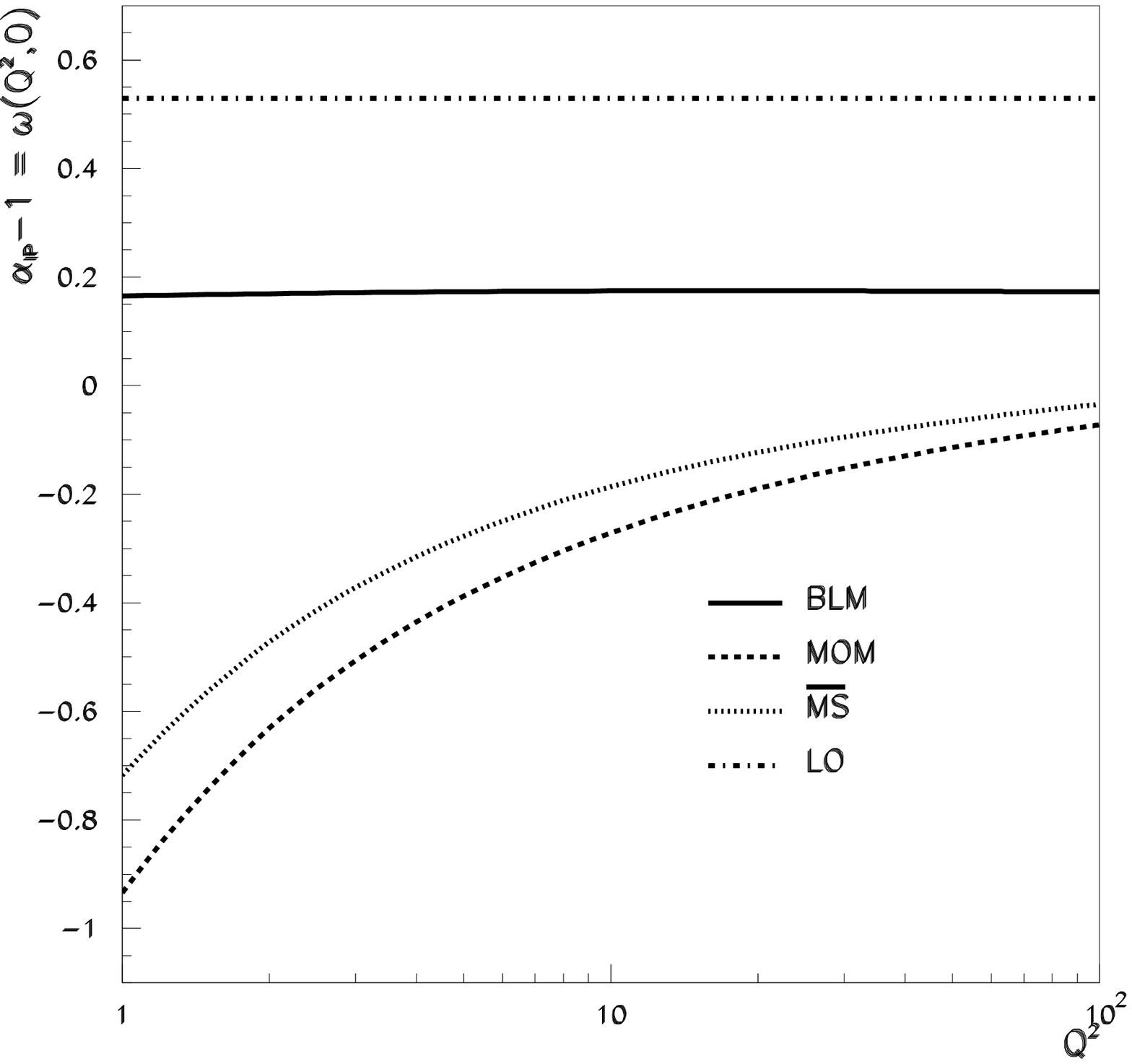}}
\end{center}
\caption[*]{$Q^2$-dependence of the BFKL Pomeron intercept in
the NLO.
The notation is as in Fig. 1.}
\label{fig:2}
\end{figure}

Figures \ref{fig:1} and \ref{fig:2} give the results for
the eigenvalue of the NLO BFKL kernel.
We have used the QCD parameter $\Lambda = 0.1$ GeV which
corresponds to $\alpha_S = 4 \pi / \bigl[ \beta_0
\ln(Q^2/\Lambda^2) \bigr]
\simeq 0.2$ at $Q^2=15$ GeV$^2$.
Also, the generalizations \cite{BGMR,Shirkov,BGKL} of the $\beta$
-function  in the running coupling and of flavor number
for continuous treatment of quark thresholds have been used.

One can see from Fig. \ref{fig:1}, that the maximum which
occurs at non-zero $\nu$ is
not as pronounced in the BLM approach compared to the
$\overline{\mbox{MS}}$-scheme, thus it does not serve
as a good saddle point at high energies.

One of the striking features of this analysis is that the NLO value for
the intercept of the BFKL Pomeron, improved by the BLM
procedure, has a
very weak dependence on the gluon virtuality $Q^2$.
This agrees with the conventional Regge theory where
one expects a universal intercept of the Pomeron without any
$Q^2$-dependence.
The minor $Q^2$-dependence obtained, on one side, provides near
insensitivity
of the results versus precise value of $\Lambda$, and, on the other
side, leads
to the approximate scale and conformal invariance. Thus one may use
conformal
symmetry \cite{Lipatov97,Lipatov86,Kotikov} for the continuation of the
present results
to the case $t \neq 0$.

Therefore, by applying the BLM scale setting
 within non-Abelian physical schemes
(MOM- and $\Upsilon$- schemes),
we do not face the serious problems \cite{Ross,Kov98,Blu98}
which were
present in the
$\overline{\mbox{MS}}$-scheme, {e.g.}, oscillatory cross
section disbehavior based on the saddle point approximation
\cite{Ross}, and a somewhat puzzling analytic structure
\cite{Kov98} of the $\overline{\mbox{MS}}$-scheme result
\cite{FL,CC98}.

 Since the BFKL equation can be interpreted as a ``quantization''
of the renormalization group
equation \cite{Lipatov86}, it follows that the effective scale
should depend on the BFKL eigenvalue $\omega$, associated with
the Lorentz spin, rather than on $\nu$.
Thus, strictly speaking, one can use the above
effective scales as function
of $\nu$ only in ``quasi-classical'' approximation at large $Q^2$.
 However, the present remarkable $Q^2$-stabilty  leads us to
expect that
the results obtained with LO eigenfunctions may not change
considerably
for $t \neq 0$ due to  the approximate conformal invariance.

\section{Other Approaches to Perturbation \newline Theory
Optimization}

Now we consider briefly the NLO BFKL within other approaches to
the
optimization of perturbative theory, namely, fast apparent
convergence (FAC)
\cite{FAC} and the principle of minimal sensitivity (PMS)
\cite{PMS}.

By the use of the FAC \cite{FAC}, one can obtain
\begin{eqnarray}
&& \omega_{FAC}(Q^2,\nu)\ =\ N_C \chi_{L} (\nu)
\frac{\alpha_{S}(Q^{2}_{FAC}(\nu))}{\pi}\ ,\\
&& Q^{2}_{FAC} (\nu)\ =\ Q^2 \exp \Biggl[-\frac4{\beta_0} r(\nu)
\Biggr].
\end{eqnarray}

In the $\overline{\mbox{MS}}$-scheme at $\nu=0$,
$\omega_{FAC}=0.33-0.26$ for $Q^2=1-100$ GeV$^2$. However,
the NLO coefficient $r(\nu)$ and hence
FAC effective scale, each have a singularity at
$\nu_0 \simeq 0.6375$ due to a zero of the $\chi_L(\nu)$-function.

In the PMS approach \cite{PMS} the NLO BFKL eigenvalue reads
as follows
\begin{equation}
\omega_{PMS}(Q^2,\nu) = N_C\,\chi_{L} (\nu)
\frac{\alpha_{PMS}(Q^{2}(\nu))}{\pi} \Biggl[ \frac{1+ (C/2)
\alpha_{PMS}/\pi } {1+ C \alpha_{PMS}/\pi} \Biggr],
\end{equation}
where the PMS effective coupling $\alpha_{PMS}$ is a solution of
the
following transcendental equation
\begin{equation}
\frac{\pi}{\alpha_{PMS}}+ C \ln \Biggl( \frac{C \alpha_{PMS}/\pi}
{ 1 + C \alpha_{PMS}/\pi} \Biggr) +
\frac{C/2}{1 + C \alpha_{PMS}/\pi} =
\frac{\beta_0}{4} \ln \Biggl( \frac{Q^2}{\Lambda^2}\Biggr) -
r(\nu),
\end{equation}
with $C=\beta_1/(4 \beta_0)$ and $\beta_1=102-38 N_F/3$. At
$\nu=0$
one obtains in the $\overline{\mbox{MS}}$-scheme
$\omega_{PMS}=0.23-0.20$ for $Q^2=1-100$ GeV$^2$
but, by the same reason as in
the FAC case, the PMS effective coupling has a singularity at
$\nu_0$.
Thus, the application of the FAC and PMS scale setting
approaches to the
BFKL eigenvalue problem leads to difficulties with the conformal
weight
dependence, which is an essential ingredient of the BFKL calculations.
The unphysical behavior of the FAC and PMS effective scales
for jet production processes has been noted in Refs. \cite{Kra91}.

The problem can be resolved by the expansion of $\chi_L(\nu)$-function
near its zero to avoid unphysical behavior
of the optimization procedure.

\section{Application for Gamma-Gamma Scattering}

The gamma-gamma total cross section calculated
with the resummation of the leading energy logarithms was considered in
\cite{BL78,Brodsky97,Bartels96}.

The total cross section of two unpolarized gammas with virtualities
$Q_{A}$
and $Q_B$ in the LO BFKL \cite{Brodsky97,BL78} reads as
follows:
\begin{equation}
\sigma(s,Q_A^2,Q_B^2)=\!\sum_{i,k = T,L}
\frac{1}{\pi Q_A Q_B} \int\limits_0^\infty
\frac{d \nu}{2 \pi}
 \cos \Biggl(\nu \ln \biggl(\frac{Q_A^2}{Q_B^2}\biggr)\Biggr)
F_{i}(\nu) F_{k}(-\nu)
\Biggl( \frac{s}{s_0}\Biggr)^{\omega(Q^2,\nu)},
\label{eqn:sigma-g}
\end{equation}
with the gamma impact factors in the LO
for the transverse and longitudinal polarizations:
\begin{eqnarray}
 F_T(\nu) = \alpha_{QED} \, \alpha_S \,
\Bigg( \sum_q e_q^2 \Bigg)  \frac{\pi}{2}
\frac{\Bigl[\frac{3}{2} - i \nu \Bigr]
\Bigl[\frac{3}{2} + i \nu \Bigr]
\Gamma \Bigg(\frac{1}{2} - i \nu \Bigg)^2
\Gamma \Bigg(\frac{1}{2} + i \nu \Bigg)^2}
{\Gamma (2 - i \nu) \Gamma (2 + i \nu)}\ ,
\label{eqn:impactt}
\end{eqnarray}

\begin{eqnarray}
 F_L(\nu) = \alpha_{QED} \, \alpha_S \,
\Bigg( \sum_q e_q^2 \Bigg) \pi
\frac{\Gamma \Big(\frac{3}{2} - i \nu \Big)
\Gamma \Big(\frac{3}{2} + i \nu \Big)
\Gamma \Big(\frac{1}{2} - i \nu \Big)
\Gamma \Big(\frac{1}{2} + i \nu \Big)}
{\Gamma (2 - i \nu) \Gamma (2 + i \nu)}\ ,
\label{eqn:impactl}
\end{eqnarray}
where Regge scale parameter $s_0$ is proportional to
a hard scale $Q^2 \sim Q_A^2,Q_B^2$, $\Gamma$ being the Euler
$\Gamma$-function, and $e_q$ is the quark electric charge.

In the NLO BFKL case one should obtain the formula analogous to
LO BFKL (Eq. \ref{eqn:sigma-g}). It has been demonstrated in
Ref. \cite{Fadin99} that the infrared singularities
at the NLO are cancelled out for impact factors of colorless
particles. Therefore, in the NLO both the kernel of the BFKL equation
and impact factors are infrared safe which confirm a
self-consistence of such factorization scheme.

While exact NLO impact factors of gamma are not known yet
\cite{Fadin:2000jj}
one can use the LO impact factors of Eqs.
(\ref{eqn:impactt}-\ref{eqn:impactl}) \cite{GLF,BL78,Brodsky97}
implying
that the main NLO corrections come from the NLO
BFKL subprocess rather than from the impact factors 
\cite{KLP99a,KLP99b}.
Thus, in the NLO BFKL one can have Eq. (\ref{eqn:sigma-g})
but with $\omega(Q^2,\nu)$ taken in the NLO.
To imply the BLM procedure to the total cross section
one can see that one can imply the BLM procedure
directly to the NLO BFKL eigenvalue $\omega(Q^2,\nu)$
within the accuracy up to the next-to-next-to-leading order
(NNLO) and higher subleading terms.

\begin{figure}[ht]
\centerline{
\epsfxsize=8cm\epsfysize=7cm\epsfbox{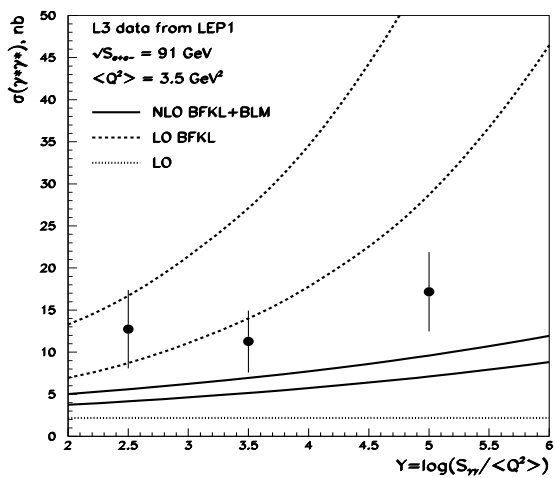}}
\caption[*]
{The NLO BFKL Pomeron vs preliminary
L3 data on virtual gamma-gamma cross section
(with subtracted quark-box contribution)
at energy 91~GeV of the $e^+e^-$ collisions.
Solid curves: NLO BFKL in BLM; dashed: LO BFKL,
and dotted: LO contribution.
Two different choices of the Regge scale: $s_0= Q^2/2$ and
$s_0=2 Q^2$.}
\label{fig:3}
\end{figure}

For numerical calculations the NLO BFKL eigenvalue
$\omega(Q^2,\nu)$ in the MOM-scheme
(Yennie gauge: $\xi=3$) has been used.

\begin{figure}[ht]
\begin{center}
\leavevmode
{\epsfxsize=8cm\epsfysize=7cm\epsfbox{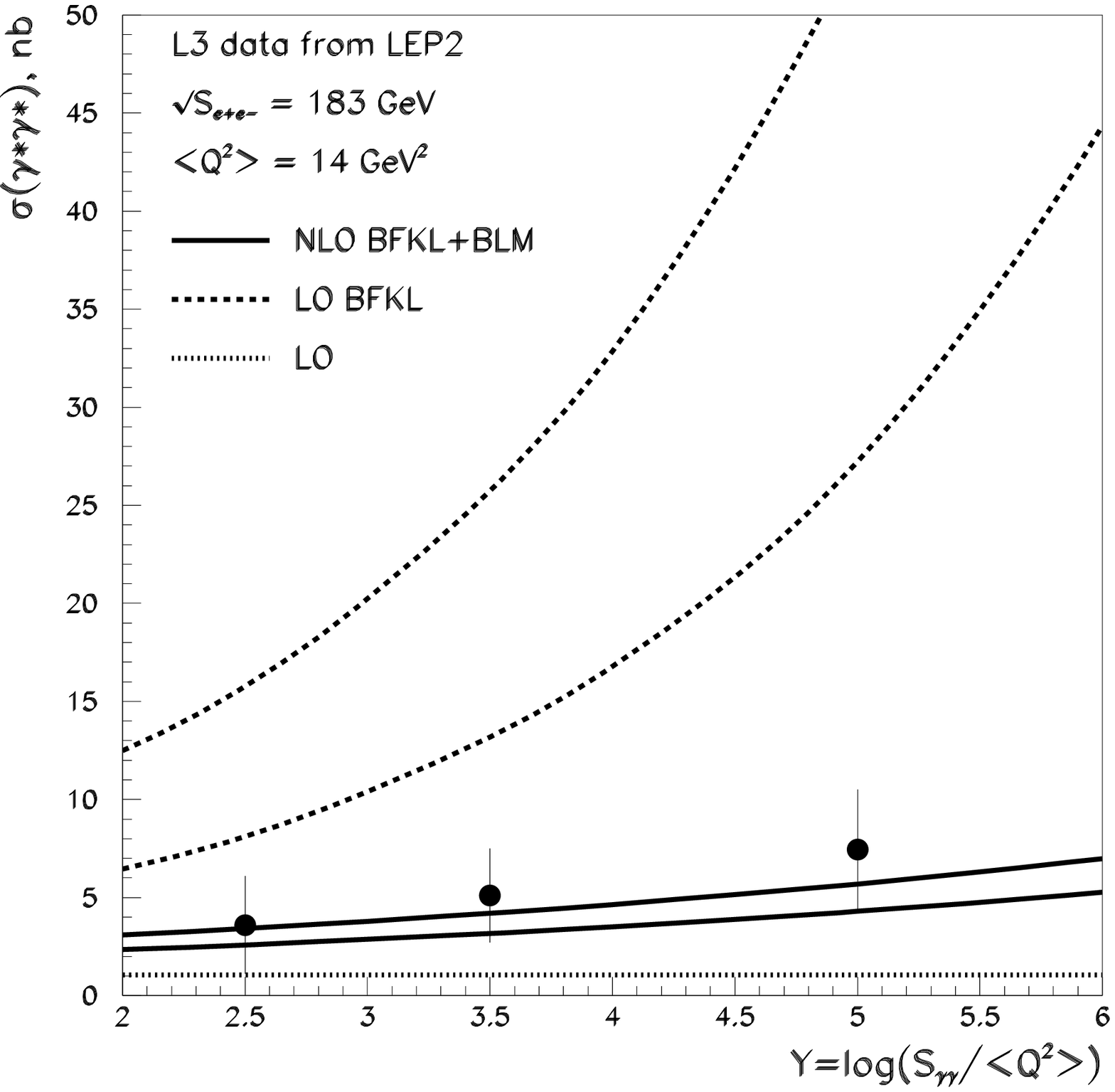}}
\caption[*]{
The same as Fig. 3, but for  energy 183~GeV
of $e^{+}e^{-}$ collisions.}
\label{fig:4}
\vspace*{0.6cm}
\leavevmode
{\epsfxsize=8cm\epsfysize=7cm\epsfbox{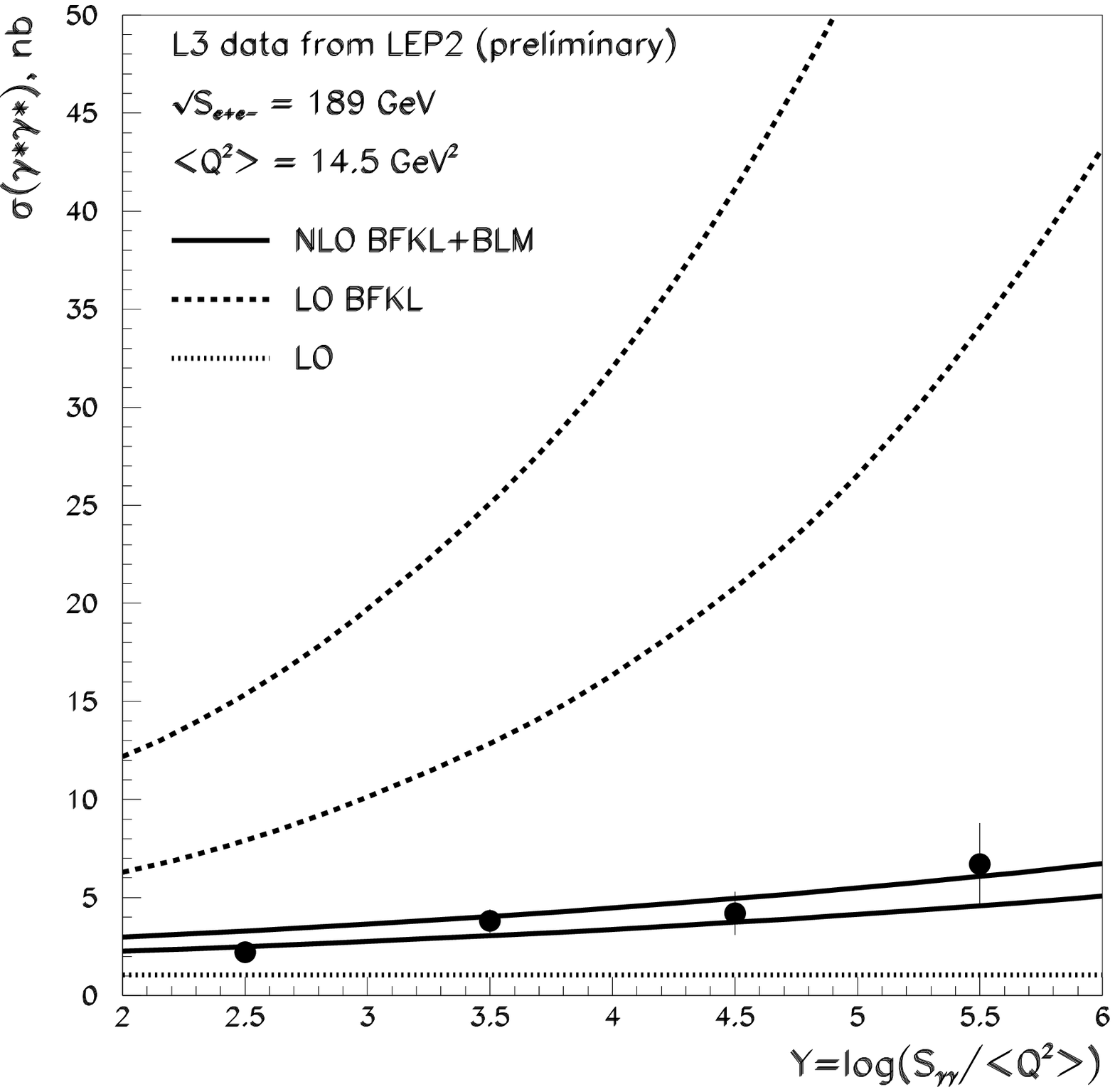}}
\end{center}
\caption[*]{
The same as Figs. 3,4, but for  energy 189 GeV
of $e^{+}e^{-}$ collisions.}
\label{fig:5}
\end{figure}

In Figs. \ref{fig:3} -- \ref{fig:5} the comparison 
of BFKL predictions in the LO and NLO
BFKL \cite{KLP99a,KLP99b} improved by 
the BLM procedure with L3 Collaboration data
\cite{L3,L3new} from
CERN LEP is shown. 
Different curves reflect uncertainty with the
choice of the Regge scale parameter which indicates when the
asymptotic
regime starts. At infinite collision energies,
the cross sections do not depend on this scale parameter $s_0$.
For present calculations, two variants have been choosen $s_0=Q^2
/ 2$
and $s_0 = 2 Q^2$, where for symmetric virtuality case
$Q^2=Q_A^2=Q_B^2$. One can see from Figs. \ref{fig:3} -- \ref{fig:5}
that the LO BFKL predictions overestimate the L3 data, while
the agreement of the NLO BFKL improved by the BLM procedure
is reasonably well, especially at higher energies of LEP2
$\sqrt{s_{e^+e^-}}=$ 183 -- 189 GeV (Figs. \ref{fig:4}, \ref{fig:5}).
One can notice also that sensitivity of the NLO BFKL results
with respect to the Regge parameter $s_0$ is much smaller than in
the case of the LO BFKL. Recent OPAL Collaboration data \cite{OPAL}
are also in a good agreement with the NLO BFKL predictions.

The gamma-gamma scattering is attractive from viewpoint
that
it is theoretically more controllable rather than hadron-hadron
and lepton-hadron collisions where non-perturbative hadronic
structure
functions are involved. In addition, in the gamma-gamma scattering
the unitarization (screening) corrections
due to multiple Pomeron exchange would be less important than in
hadron collisions. It was shown in Refs. \cite{Kaidalov86}
that the unitarization corrections in hadron collisions
can lead to higher value of the (bare) Pomeron intercept than
the effective intercept value. Since the hadronic data fit yields
about 1.1 for the effective intercept value \cite{Cudell99,Cudell97},
then the bare Pomeron intercept value should be above this value.
So that, in case of small unitarization corrections in the gamma-
gamma
scattering at large $Q^2$ one can accomodate the NLO BFKL
Pomeron
intercept value 1.13-1.18 along with larger unitarization corrections
in hadronic scattering \cite{Kaidalov86}, where it can lead to
a smaller effective Pomeron intercept value about 1.1 for
hadronic collisions.

\section{Summary}

There have been a number of recent papers which analyze the NLO
BFKL
predictions in terms of  rapidity correlations
\cite{Lipatov98,Schmidt99}, $t$-channel unitarity \cite{Coriano95},
angle-ordering \cite{CCFM},  double transverse momentum
logarithms \cite{Andersson96,Salam98,Ciafaloni99,Ball99} and  BLM
scale setting
for deep inelastic structure functions  \cite{Thorne99}.
This requires a further study to find relations between such
approaches.

To summarize, we have shown that the NLO corrections to the
BFKL
equation for the QCD Pomeron become controllable
and meaningful provided one uses physical renormalization
schemes relevant to non-Abelian gauge theory.  BLM optimal scale
setting
automatically sets the appropriate physical renormalization scale
by
absorbing the non-conformal $\beta$-dependent coefficients.   The
strong renormalization scale dependence of the NLO
corrections to BFKL resummation then largely disappears.
This is in contrast to the unstable
NLO results obtained in the conventional $\overline{\mbox{MS}}$-
scheme with
arbitrary choice of renormalization scale.
A striking feature of the NLO BFKL Pomeron intercept in
the BLM approach is its very weak $Q^2$-dependence, which
provides
approximate conformal invariance. The NLO BFKL application
to the total gamma-gamma cross section shows a good agreement
with the preliminary L3 data at the CERN LEP2 energies.
The results presented here open new windows for applications
of NLO BFKL resummation to the high-energy phenomenology.

{\bf Acknowledgements.}
We would like to thank S.~J.~Brodsky for
fruitful collaboration and L.~G.~Dakhno,
J.~R.~Ellis, A.~B.~Kaidalov,
A.~L.~Kataev, V.~A.~Khoze,
A.~V.~Kotikov, E.~A.~Kuraev and J.~P.~Vary
for helpful conversations. We thank V.~P.~Andreev, A.~De Roeck,
M.~N.~Kienzle-Focacci, C.-N.~Lin, V.~A.~Schegelsky,
A.~A.~Vorobyov and M.~Wadhwa for
useful discussions concerning the CERN LEP data.
This work was supported in part by the Russian
Foundation for Basic Research (RFBR), the INTAS foundation,
 and the U. S. National Science Foundation.

{\it Note added at Proof.} After presentation of this lecture,
the OPAL and L3 Collaborations
at the CERN LEP accumulated statistics at higher energies
and finalized their data \cite{OPALL3}.
The final OPAL and L3 data  \cite{OPALL3}, although presented
in a different way, show even better agreement
\cite{BFKLP2}
with our earlier predictions \cite{KLP99a,KLP99b}.


\end{document}